%Paper: hep-ph/9403260
%From: ereidell@marie.mit.edu (Evan Reidell)
%Date: Wed, 9 Mar 1994 17:58:00 -0500

%%%
%%%
%%%
\vsize=8.5truein
\hsize=6.0truein
\hoffset=.25in
\voffset=.25in
\input epsf

\tolerance 500
\hfuzz=50pt

%%% This twelvepoint math font family is needed
%%% for the World Scientific layout requirements
%%% TeXbook p.350 (where the ten family gets defined)
%%% MAKE SURE YOUR 12pt IS IN THE STANDARD DISTRIBUTION!

\font\twelverm = cmr12
\font\twelvei  = cmmi12
\font\twelvesy = cmsy10 at 12pt %%% scaled\magstep1
\font\twelvebf = cmbx12
\font\twelveit = cmti12
\font\twelvesl = cmsl12
\font\twelvett = cmtt12

\font\tenrm = cmr10
\font\tenbf = cmbx10
\font\tenex = cmex10

\font\ninerm =  cmr9
\font\ninei  = cmmi9
\font\ninesy = cmsy9
\font\ninebf = cmbx9

\font\sixrm=cmr6
\font\sixi =cmmi6
\font\sixsy=cmsy6
\font\sixbf=cmbx6

%%% TeXbook p.414 (where the tenpoint family gets defined)

\def\twelvept{\def\rm{\fam0\twelverm}% switch to 12-point type
  \textfont0=\twelverm   \scriptfont0=\ninerm    \scriptscriptfont0=\sixrm
  \textfont1=\twelvei    \scriptfont1=\ninei     \scriptscriptfont1=\sixi
  \textfont2=\twelvesy   \scriptfont2=\ninesy    \scriptscriptfont2=\sixsy
  \textfont3=\tenex      \scriptfont3=\tenex     \scriptscriptfont3=\tenex
  \textfont\bffam=\twelvebf   \scriptfont\bffam=\ninebf
  \scriptscriptfont\bffam=\sixbf  \def\bf{\fam\bffam\twelvebf}
  \textfont\itfam=\twelveit       \def\it{\fam\itfam\twelveit}
  \textfont\slfam=\twelvesl       \def\sl{\fam\slfam\twelvesl}
  \textfont\ttfam=\twelvett       \def\tt{\fam\ttfam\twelvett}
  \normalbaselineskip=16pt plus 1pt
  \setbox\strutbox=\hbox{\vrule height11pt depth5pt width0pt}
  \let\sc=\ninerm
  \normalbaselines\rm}

\def\footnoterule{\kern-3pt \hrule width \hsize \kern2.6pt}
\def\square{\kern1pt\vbox{\hrule height1.2pt\hbox{\vrule width1.2pt\hskip3pt
\vbox{\vskip 6pt}\hskip 3pt\vrule width.6pt}\hrule height.6pt}\kern1pt}

\baselineskip 12pt plus 1pt minus 1pt
\vskip 12pt
\centerline{{\twelvebf
$Q^2$-DEPENDENCE OF DEEP INELASTIC SUM RULES}\footnote{*}
{\ninerm This work is supported in part by funds provided by the U. S.
Department of Energy (D.O.E.) under contract \#DE-AC02-76ER03069.}}
\vskip 20pt
\centerline{XIANGDONG JI}
\vskip 5pt
\centerline{\it Center for Theoretical Physics}
\centerline{\it Laboratory for Nuclear Science}
\centerline{\it and Department of Physics}
\centerline{\it Massachusetts Institute of Technology}
\centerline{\it Cambridge, Massachusetts\ \ 02139\ \ \ U.S.A.}
\vskip .4in

%% \centerline{ABSTRACT}
%% \smallskip
%% {\narrower
%% Perhaps the introduction below can be reworded to make an abstract.
%% \smallskip}

\twelvept
\baselineskip 14pt plus .5pt minus .5pt

In this talk, I will concentrate on $Q^2$-dependence of deep inelastic sum
rules.  I will first give a modern definition of deep-inelastic sum rules and
then discuss physical origins of their scaling violation at finite $Q^2$.
Following this, I discuss a few well-known examples, in particular, the
Bjorken sum rule, which is at the center of interest of this symposium.
Finally, I consider the $Q^2 \to 0$ limit of sum rules using low-energy
theorems.  I think this can motivate some interesting CEBAF physics.

\goodbreak
\vskip16pt
\line {\bf 1.\quad Deep-Inelastic Sum Rules \hfil}
\vskip 8pt
\nobreak
Let me start with the structure functions of the nucleon.   In inclusive
lepton-nucleon scattering, one measures the following hadron tensor,
$$
W_{\mu\nu} = {1\over 4\pi} \int e^{i q \cdot \xi} \, d^4\xi
\, \langle \, PS \, | \, J_\mu^{\dagger}(\xi) \, J_\nu(0) \,
| \, PS \, \rangle ~~,
\eqno(1)
$$
where $J_\mu$ is the electroweak current of quarks, $S$ and $P$ are
polarization and momentum vectors of the nucleon, respectively, and $q$ is the
momentum of a virtual boson.  $W_{\mu\nu}$ can be decomposed into various
scalar structure functions of the nucleon$^1$, which are shown in Table~1.
The different column shows dependence on the nucleon polarization:
Unpolarized, longitudinally-polarized, and transversely polarized.  The
parity-even structure functions are measurable in electromagnetic processes,
whereas the parity-odd ones couple through weak interactions.  The structure
functions
with circles are the ones that have been measured in previous experiments.
The cross or check under each structure function labels the beam polarization
(no or yes) when it is measured.


\newdimen\totalht

\def\circled#1{%
\setbox0=\hbox{$#1$} \totalht=\ht0 \advance\totalht by -\dp0
\kern0.5\wd0 \raise0.5\totalht
\hbox to 0pt{
}
\kern-.5\wd0
#1}

%%
%% you've got to!  literal postscript circles, shaded even!!
%%
\midinsert
\medskip
\centerline{\tenrm {\tenbf Table I}: Structure functions}
\medskip
\hbox to \hsize{\hfil\vbox{
\halign{\vrule width0pt height 15pt depth 7pt
\enskip\hfil#\hfil\enskip
&\enskip\hfil#\hfil\enskip
&\enskip\hfil#\hfil\enskip
&\enskip\hfil#\hfil\enskip\cr
\noalign{\hrule}
{~~} & Un-polarized & L-polarized & T-polarized \cr
\noalign{\hrule}
$P$-even &
$\circled{W_{\!2}} \,$, ~ $\circled{W_{\!L}}$ & $\circled{G_1}$ & $G_2$ \cr
\noalign{\vskip-4pt}
(Beam-pol) & $\times$ & $\surd$ & $\surd$ \cr
\noalign{\hrule}
$P$-odd & $\circled{W_{\!3}}$ & $X_L \,$, ~ $X_2$ & $Y_1$ \cr
\noalign{\vskip-4pt}
(Beam-pol) & $\surd$ & $\times$ & $\times$ \cr
\noalign{\hrule}}}\hfil}
\medskip
\endinsert

The structure functions
depend on two Lorentz scalars,
$Q^2 = -q^2$ and $\nu = p \cdot q$.
In the deep-inelastic limit,
{\it i.e.\/},
$Q^2,\nu \to \infty$,
$Q^2 / 2 \nu = x = \, {\rm fixed}$,
the structure functions scale to quark distributions,
which are only functions of $x$,
(neglecting renormalizations point dependence for the moment),
$$
\eqalign{
W_2 &\to F_2 \sim q(x) + \bar{q}(x) ~~, \cr
W_3 &\to F_3 \sim q(x) - \bar{q}(x) ~~, \cr
G_1 &\to g_1 \sim \Delta q(x) + \Delta\bar{q}(x) ~~, \cr
X_1 &\to a_1 \sim \Delta q(x) - \Delta\bar{q}(x) ~~, \cr}
\eqno(2)
$$
where quark distribution $q(x)$ and quark helicity distribution $\Delta q(x)$
are defined as,
$$
\eqalign{
q(x) &= \int {d \lambda \over 2\pi} \, e^{i\lambda x} \,
\langle \, P \, | \, \bar{\psi} (0) \, n \!\!\! / \, \psi(\lambda n) \,
| \, P \, \rangle ~~, \cr
\Delta q(x) &= \int {d \lambda \over 2\pi} \, e^{i\lambda x} \,
\langle \, PS \, | \, \bar{\psi} (0) \, n \!\!\! / \, \gamma_5 \,
\psi(\lambda n) \, | \, PS \, \rangle ~~, \cr}
\eqno(3)
$$
where $n$ is a light-like vector.  Two things can be said about these
distributions.  First, they are light-cone correlations as the quark fields
in the matrix elements are separated along the light-cone.   Second, they are
related to the ground state properties of the nucleon.  Because of this second
property, one can immediately derive the structure function sum rules at $Q^2
= \infty$ (the deep inelastic sum rules),
$$
%%
%% remember that stupid bar on the lecture notes!
%%
\eqalign{
\int\limits_0^1 F_3^{\nu p+\bar{\nu}p} (x) \, dx &= 3 ~~, \cr
\int\limits_0^1 g_1^{e p - e n} (x) \, dx &= {1\over6} g_A ~~, \cr
{18\over5} \int\limits_0^1 F_2^{ep+en}(x) \, dx &= \Delta p ~~. \cr}
\eqno(4)
$$
The first is the Gross-Llewellyn Smith sum rule, the second is the Bjorken sum
rule, and the third is the momentum sum rule with $\Delta p$ denoting the
momentum fraction of the nucleon carried by quarks.

\goodbreak
\vskip16pt
\line {\bf 2.\quad Scaling Violation At Finite $Q^2$ \hfil}
\vskip 8pt
\nobreak
At finite $Q^2$, the deep-inelastic sum rules are violated by quark-gluon
interactions.  Thus one can write a generalized sum rule,
$$
\int\limits_0^1 ( \ldots ) \, dx = \Gamma(Q^2) ~~,
\eqno(5)
$$
where $\Gamma(Q^2)$ is an unknown function of $Q^2$, except at the $Q^2 \to
\infty$ limit, where it approaches the deep-inelastic limit $\Gamma_0$.  A
schematic drawing for $\Gamma(Q^2)$ is shown in Fig.~1.

\midinsert
\hbox to \hsize{\hss\epsffile[32 52 335 178]{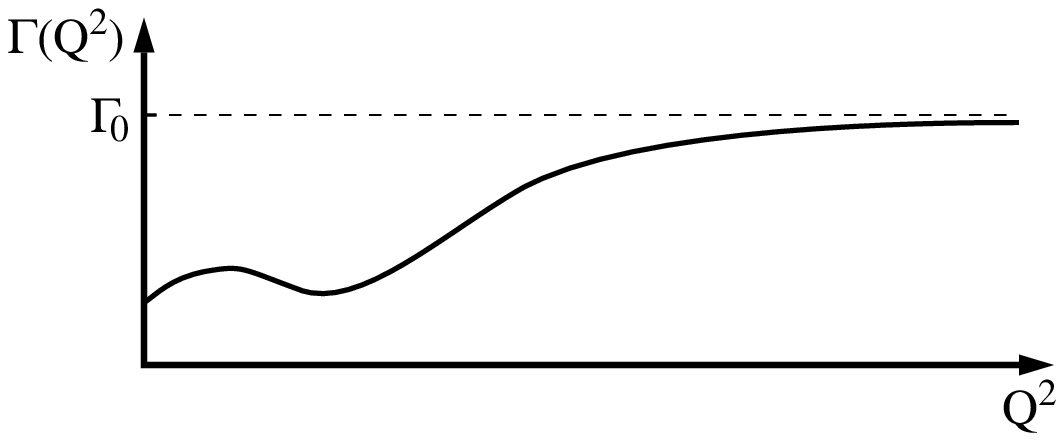}\hskip40pt\hss}
\medskip
\centerline{\tenrm {\tenbf Figure~1}.
Schematic $Q^2$-dependence of a deep inelastic sum rule.}
\endinsert

At present, there is no general theory about $Q^2$ variation of $\Gamma(Q^2)$.
However, for $Q^2 > Q_0^{\,2}$, some scale related to hadron masses or
non-preturbative physics, we believe the following twist expansion is correct,
$$
\Gamma(Q^2) = E_0 \left( {Q^2 \over \mu^2} \right) \Gamma_0(\mu^2)
+ E_2 \left( {Q^2 \over \mu^2} \right) \Gamma_2 (\mu^2) \Bigl/ Q^2
+ E_4 \left( {Q^2 \over \mu^2} \right) \Gamma_4 (\mu^2) \Bigl/ Q^4 +
\cdots ~~,~
\eqno(6)
$$
where the coefficient functions have expansions in the strong coupling
constant,
$$
E_n = \sum\limits_{i=0}^\infty \alpha_s^{~i} (Q^2) e_n^i ~~.
\eqno(7)
$$
The terms beyond the first in Eq.~(6) are suppressed by successive powers of
$1/Q^2$ and are called higher twist corrections.  $\Gamma_n(\mu^2)$ are related
to the nucleon matrix elements of {\it local\/} operators containing quark and
gluon fields,
$$
\Gamma_n (\mu^2) = \langle \, PS \, | \, \hat{O}_n (\mu^2) \, | PS \, \rangle
{}~~.
\eqno(8)
$$

It is clear, then, that $Q^2$ dependence of $\Gamma(Q^2)$ is introduced
through both power-dependence (higher twists) and logarithmic dependence in
the strong coupling constant (QCD radiative corrections).  Both contributions
can be represented by Feynman diagrams as shown in Fig.~2.

\midinsert
\hbox to \hsize{\hss\epsffile[50 43 508 160]{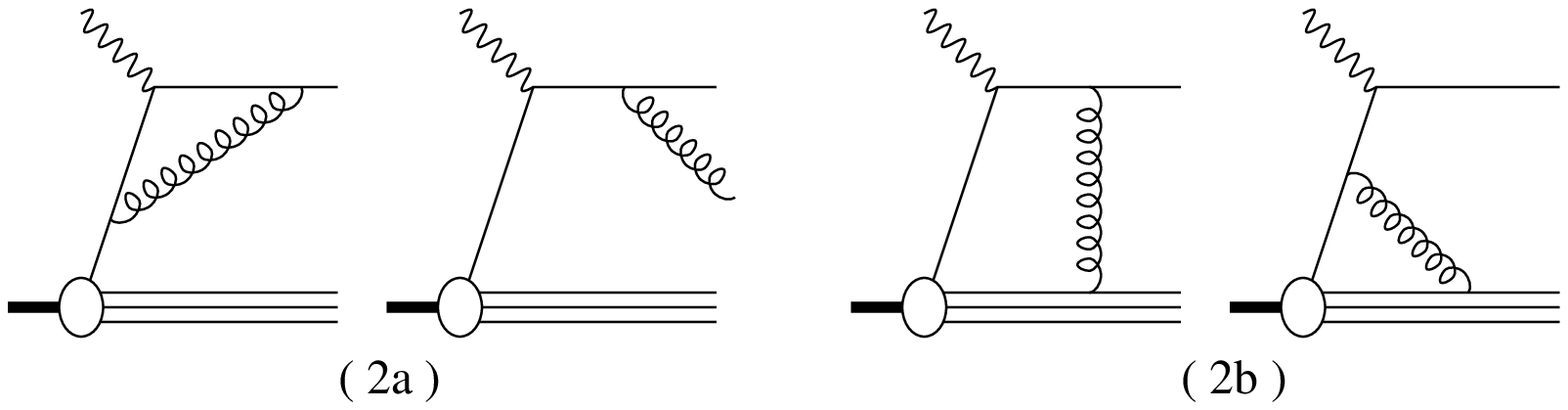}\hss}
\advance\leftskip by 0.8in
\advance\rightskip by 0.8in
\medskip\noindent
{\tenrm {\tenbf Figure~2}.
Feynman diagrams giving $Q^2$-dependence to deep inelastic
sum rules: (2a) radiative corrections, (2b) higher-twist effects.}
\endinsert

It must be emphasized, however, the above picture is not complete.  Recently,
works were produced which go beyond this canonical understanding.  Some
examples: Balitsky and Braun have considered instanton contributions to
deep-inelastic structure functions, which cannot simply be classified as power
or log corrections$^2$; Muller has considered non-perturbative improvement to
the perturbative series in (7) and has shown that it gives rise to power-like
terms which are not included in (6) [Ref.~3].  In the subsequent discussion, I
assume these contributions are small.

\goodbreak
\vskip16pt
\line {\bf 3.\quad Examples \hfil}
\vskip 8pt
\nobreak
In the following discussion, I show
three examples of twist expansion:  the Bjorken sum rule, the
Gross-Llewellyn Smith sum rule, and the $F_L(x)$ sum rule.

With $Q^2$-dependent corrections, the Bjorken sum rule can be written as,
$$
\int\limits_0^1 g_1^{p-n}(x) \, dx = {g_A \over 6}
\left(
1 - {\alpha_S \over \pi} - 3.58 \left( \alpha_S \over \pi \right)^2
- 20.2 \left( {\alpha_S \over \pi} \right)^3 + \cdots
\right)
+ {\mu_4^{~p-n} \over Q^2} + \cdots
\eqno(9)
$$
where QCD radiative corrections to the twist-two contribution have been
calculated by Larin {\it et al.}$^4$.
The first non-trivial power corrections were first calculated by Shuryak and
Vainshtein$^5$, and have recently been checked by Ji and Unrau$^6$.
The result for
$\mu_4$ is,
$$
\mu_4 = {M^2 \over 9} \sum_f e_f^{~2}
\left(
a_{2f} + 4 d_{2f} - 4 f_f
\right) ~~,
\eqno(10)
$$
where $a_{2f}$, $d_{2f}$, and $f_f$ are related to the matrix elements of
twist-two, three, and four operators.
These higher-twist matrix elements have been evaluated in terms of the QCD
sum rule method and Bag model.  (I will neglect the QCD radiative corrections
associated with these matrix elements).

Let me consider these corrections at $Q^2 = 2 \, {\rm GeV}^2$.
To evaluate the QCD radiative corrections, I take
$\Lambda^{(4)}_{\overline{MS}} = 260^{+54}_{-46}$ MeV from the particle data
book, which corresponds to $\alpha_S(M_Z) = 0.1134 \pm 0.0035$, a
world-average
including LEP data and deep inelastic scattering fits.  The error is
enlarged to 10\% at $Q^2 = 2 \, {\rm GeV}^2$,
$$
\alpha_S(2 \, {\rm GeV}^2) = 0.330 \pm 0.035  ~~.
\eqno(11)
$$
This gives the number in the bracket in (9) $0.835^{+0.023}_{-0.031}$,
representing a 17\% correction.  The correction from still higher power of
$\alpha_S$ terms is about $20 \left( {\alpha_S \over \pi} \right) \sim 0.03$,
a 3\% correction.  Thus, neglecting the higher twists, the Bjorken sum rule at
$Q^2 = 2 \, {\rm GeV}^2$ is
$$
\int\limits_0^1 g_1^{~p-n} (x,Q^2) \, dx = 0.175 \pm 0.008 ~~.
\eqno(12)
$$
The higher twist matrix elements have been evaluated in the Bag model by Ji
and Unrau$^6$, who found,
$$
{M_4^{~p-n} \over Q^2} = 0.031 {M^2 \over Q^2} = 0.014 ~~,
\eqno(13)
$$
which is an 8\% correction.  On the other hand, the same matrix elements, when
evaluated in QCD sum rule (Balitsky, Braun, Kolesnichenko$^7$), give,
$$
{M_4^{~p-n} \over Q^2} = -0.023 {M^2 \over Q^2} = -0.011 ~~,
\eqno(14)
$$
which is a 6\% correction with an opposite sign to (13).  Given this large
uncertainty on the matrix elements, I give a $\pm 0.014$ correction to the
Bjorken sum rule from the higher twists.  Thus, the best knowledge for the sum
rule from the theoretical side is,
$$
\int\limits_0^1 g_1^{~p-n} (x) \, dx = 0.175 \pm 0.008 \pm 0.014 ~~.
\eqno(15)
$$
This is consistent with the present experimental determination of this sum
rule from E142 experiment$^8$.

On the other hand, if one considers the Bjorken sum rule at
$Q^2 = 10 \, {\rm GeV}^2$,
the theoretical uncertainty is much smaller, in fact,
$$
\int\limits_0^1 g_1^{~p-n} (x) \, dx = 0.188 \pm 0.003 \pm 0.003 ~~,
\eqno(16)
$$
at $Q^2 = 10 \, {\rm GeV}^2$.

The deviation from Gross-Llewellyn smith sum rule is usually characterized by
$\Delta$ defined as,
$$
\eqalign{
\Delta &= 1 - {1\over3} I_{\rm GLS} ~~, \cr
I_{\rm GLS} &= \int_0^1 F_3^{\nu p + \bar{\nu} p} (x) \, dx ~~. \cr}
\eqno(17)
$$
Experimentally, $\Delta$ has been measured by CCFR collaboration at $Q^2 = 3
\, {\rm GeV}^2$ [Ref.~9],
$$
\Delta^{\rm exp} = 0.167 \pm 0.006 \pm 0.026 ~~.
\eqno(18)
$$
Theoretically, $\Delta$ can be expressed as
$$
\Delta^{\rm th} = Q {\alpha_S \over \pi} + 3.58 \left( {\alpha_S \over \pi}
\right)^2 + 19.0 \left( {\alpha_S \over \pi} \right)^3 + \cdots
\, + {8\over27} {d \over Q^2} + {{\rm T.M.}\over Q^2} + \cdots ~~,
\eqno(19)
$$
where $d$ is the matrix element of a twist-four operator and T.M.\ means
target mass correction.  The twist-four matrix element has been calculated in
the Bag model and in the QCD sum rule, both giving a consistent result, 0.33
GeV$^2$.  Substituting it to (19), I have
$$
\Delta^{\rm th} = 0.170 + {{\rm T.M.}\over Q^2} ~~.
\eqno(20)
$$
Without the target mass correction, $\Delta^{\rm th}$ agrees with $\Delta^{\rm
exp}$ perfectly.  However, when it is included, we have, $\Delta^{\rm th} =
0.140$, which is still within the experimental error bar.  Thus,
a better precision is needed for $\Delta^{\rm exp}$ to discern a
non-trivial higher twist effect.

Finally, I consider the ${1/Q^2}$ correction to $F_L(x)$.  It is known
that there are contributions from target masses and QCD radiative corrections
to
$F_L(x)$.  However, these contributions are not enough to explain the data on
$F_L(x)$.  The residue must come from the non-trivial higher twist
contributions.  Recently, Choi {\it et al.\/}$^{10}$
have extracted a moment of the residual
$F_L(x)$ (I call $\Delta F_L(x)$) from the SLAC-MIT, BCDMS, and NMC data,
$$
2\int_0^1 x \, \Delta F_L(x) \, dx =
\cases {
(0.035 \pm 0.012)/Q^2 & (proton) \cr
(0.023 \pm 0.008)/Q^2 & (neutron) \cr}
\eqno(21)
$$
I have calculated the higher twist contribution from the Bag model,
the result is,
$$
2\int_0^1 x \, \Delta F_L (x) \, dx =
\cases {
 0.020 / Q^2 & (proton) \cr
 0.013 / Q^2 & (neutron) \cr
}
\eqno(22)
$$
which is roughly consistent with the data.  Notice that the data shows a
remarkable $SU(6)$ structure when a ratio is made between the proton and
neutron results, which I think is a srong support for the Bag
calculation.

\goodbreak
\vskip16pt
\line {\bf 4.\quad Sum Rules At $Q^2\rightarrow 0$ \hfil}
\vskip 8pt
\nobreak
Finally, I discuss the $Q^2\rightarrow 0$ limit of the sum rules.
I will take the Bjorken sum rule as an
example although the discussion can be extended to other sum rules.
I introduce,
$$
\eqalign{
\Gamma(Q^2) &= \int\limits_0^1 g_1 (x,Q^2) \, dx \cr
&= {Q^2 \over 2M^2} \int\limits_{Q^2/2}^\infty G_1 (\nu,Q^2) {d\nu\over\nu}\cr
&= {Q^2 \over 2M^2} I_1 (Q^2) }
\eqno(23)
$$
It was observed by Ansemlino {\it et al.\/}$^{11}$
that $I_1(0)={-\kappa^2 / 4}$
from Drell-Hearn-Gerasimov sum rule, where $\kappa$ is the anomalous
magnetic moment.  Since $I_1(Q^2)$ is positive at large $Q^2$, they argue
that higher twists must be unusually large so as to make the
variation of $I_1(Q^2)$ smooth.  I will show below that this
is not the case.

What they have overlooked is the nucleon's elastic contribution to
$g_1(x,Q^2)$, which is non-analytic as function of $Q^2$.  At
$Q^2=0$, the elastic contribution vanishes because of energy
momentum conservation.  At $Q^2\not= 0$, however small it may be, the
elastic contribution to $g_1(x,Q^2)$ exists.  To look at the overall
$Q^2$ dependence of $\Gamma(Q^2)$ to draw conclusions about the higher
twists, one must not neglect this elastic contribution at low $Q^2$.
In duality language, which is implicitly assumed here, it is the
sum of elastic plus resonance contributions that duals the deep-inelastic
twist expansion.  According to these arguments, I have shown at low $Q^2$
[Ref.~12] that,
$$
\Gamma(Q^2) = {1\over2} F_1 (F_1 + F_2) - {F_2^{\,2}  \over 8 M^2} Q^2
\eqno (24)
$$
where $F_1$ and $F_2$ are Dirac and Pauli form factors.  The first term
in (24) comes from elastic contribution and the second from inelastic
contributions as summed by Drell-Hearn-Gerasimov sum rule.  Clearly, if
the nucleon is a point-like object, we have $\Gamma_1(Q^2)={1\over 2}$
at all $Q^2$.

Eq.~(24) tells us both $\Gamma(Q^2)$ and its first derivative at
$Q^2=0$.  In fact,
$$
\eqalignno{
\Gamma(0) &= \cases
{1.396 & (proton) \cr
0 & (neutron) \cr}
& (25) \cr
\left. {d\Gamma \over d Q^2} \right|_{Q^2 = 0} &=
\cases {
-8.631 {\rm GeV}^{-2} & (proton) \cr
-0.479 {\rm GeV}^{-2} & (neutron) \cr}
& (26) \cr}
$$
\midinsert
\epsfxsize=4.5truein
\hbox to \hsize{\hss\epsffile[125 120 517 428]{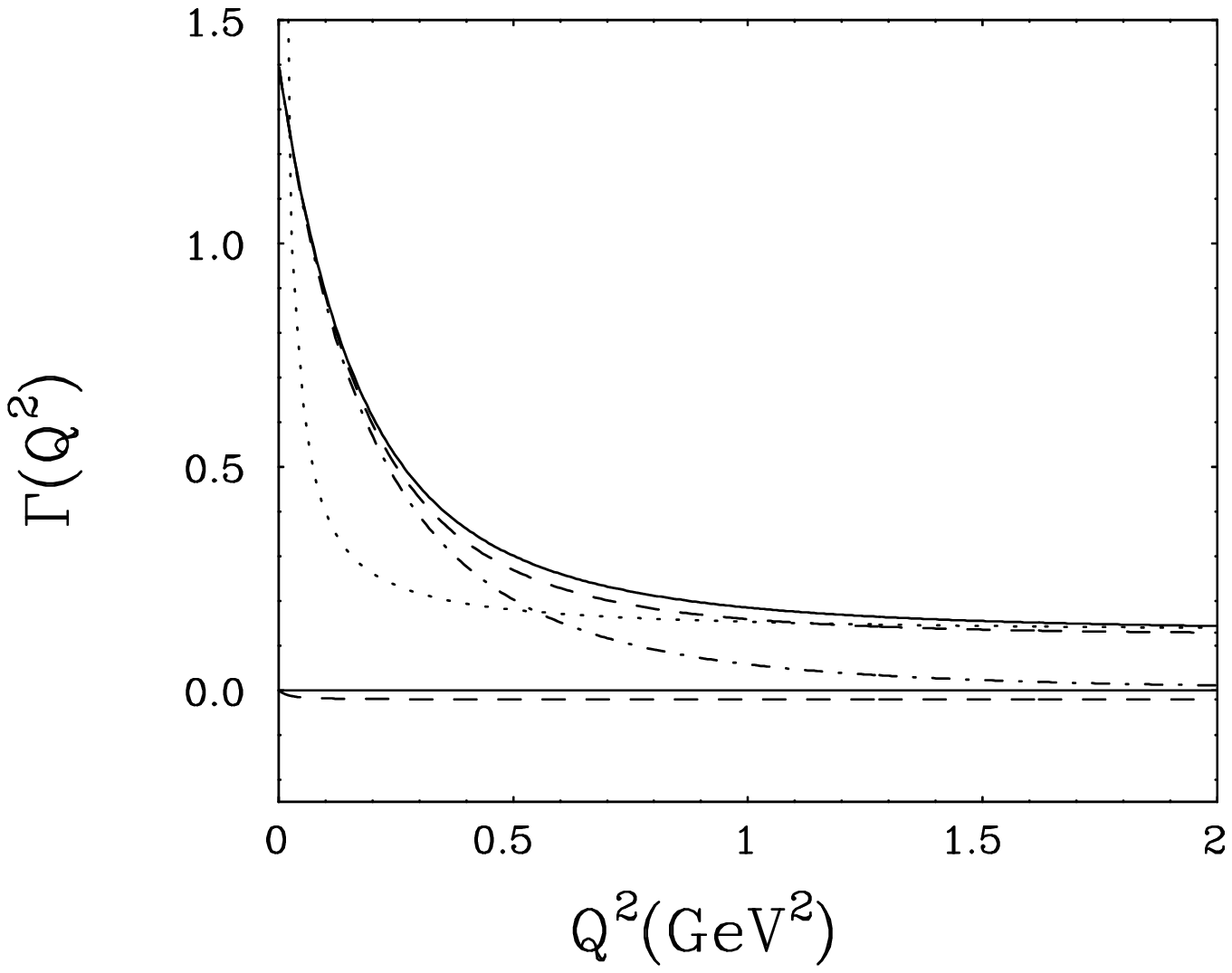}\hskip.75in\hss}
\medskip
\noindent
{\tenrm {\tenbf Figure~3}.~
A model for the sum rule $\Gamma(Q^2)$ at all $Q^2$.
The solid and upper-dashed curves are the parameterization with the bag and
QSR higher twist matrix elements, respectively.  The dotted curve represents
the result of the twist expansion to order $1/Q^2$ and the dot-dashed curve
represents the elastic contribution.  A similar interpolation for the neutron
is shown as the lower-dashed curve.}
\medskip
\endinsert
Using these, and the twist-expansion at large $Q^2$, we can construct an
interpolating sum rule,
$$
\Gamma(Q^2) = {1\over2} F_1 (F_1 + F_2)
\left( 1 - \lambda_1 {Q^2 \over M^2} \right)
+ \lambda_2 {1+\lambda_3 M^2 / Q^2 \over 1 + \lambda_4 M^4 / Q^4}
\eqno (27)
$$
which is shown in Fig.~3 for two different choices of higher twist
matrix elements.  As is clearly seen from the figure, the higher twists
of sizes from the Bag or QCD sum rule calculations are
consistent with low $Q^2$ behavior
due to the large negative derivative of $\Gamma$ at the origin.

Finally, let me comment that future experimental data on $\Gamma(Q^2)$ from
low $Q^2$ (say 0.5 ${\rm GeV}^2$) will be very useful for confirming the
picture presented above.  The data for the large $x (>0.1)$ region may be
obtained from CEBAF where resonance physics is important.

%\goodbreak
%\bigskip
%\centerline{\bf ACKNOWLEDGEMENTS}
%\medskip
%\nobreak
%\vfill\eject

\goodbreak
\bigskip
\centerline{\bf REFERENCES}
\medskip
\item{1.}
X.~Ji, {\it Nucl.~Phys.\/}~{\bf B402} (1993) 217.

\medskip
\item{2.}
I.~Balitsky and V.~Braun, {\it Phys.~Lett.\/}~{\bf B134} (1993) 237.

\medskip
\item{3.}
A.~Muller, {\it Phys.~Lett.\/}~{\bf B308} (1993) 355.

\medskip
\item{4.}
S.~A.~Larin, F.~V.~Tkachev, and J.~A.~M.~Vermaseren,
{\it Phys.~Rev.~Lett.\/}~{\bf 66} (1991) 862.

\medskip
\item{5.}
E.~Shuryak and A.~Vainshtein,
{\it Nucl.~Phys.\/}~{\bf B199} (1982) 451;
{\it Nucl.~Phys.\/}~{\bf B201} (1982) 141.

\medskip
\item{6.}
X.~Ji and P.~Unrau, MIT CTP preprint 2232, 1993.

\medskip
\item{7.}
I.~I.~Balitsky, V.~M.~Braun, and A.~V.~Kolesnichenko,
{\it JETP Lett.\/}~{\bf 50} (1989) 61; Erratum, hep-ph/9310316, 1993.

\medskip
\item{8.}
The {\bf E142} Collaboration, P.~L.~Anthony {\it et~al.\/},
{\it Phys.~Rev.~Lett.\/}~{\bf 71} (1993) 959.

\medskip
\item{9.}
The {\bf CCFR} Collaboration, W.~Leung {\it et al.\/},
{\it Phys.~Lett.\/}~{\bf B317} (1993) 655.

\medskip
\item{10.}
S.~Choi, T.~Hatsuda, Y.~Koike and S.~H.~Lee,
MSUCL-880, DOE/ER/40427-05-N93, 1993.

\medskip
\item{11.}
M.~Anselmino, B.~L.~Ioffe, and E.~Leader,
{\it Sov.~J.~Nucl.~Phys.\/}~{\bf 49} (1989) 136.

\medskip
\item{12.}
X.~Ji, {\it Phys.~Lett.\/}~{\bf 309B} (1993) 187.

\bye